\documentclass[prl,aps,twocolumn,showpacs,nobibnotes,epsf]{revtex4}

\usepackage{graphicx}
\usepackage{dcolumn}
\usepackage{bm}
\usepackage{SIunits}
\usepackage{verbatim}
\usepackage{placeins}
\usepackage{multirow}

\begin{document}
\title{ BaFe$_{2}$As$_{2}$ Surface Domains and Domain Walls: Mirroring the Bulk Spin Structure}

\author{Guorong Li$^1$, Xiaobo He$^1$, Ang Li$^2$, Shuheng  H. Pan$^2$, Jiandi Zhang$^1$, Rongying Jin$^1$, A. S. Sefat$^3$, M. A. McGuire$^3$, D. G.
Mandrus$^{3, 4}$, B. C. Sales$^3$, and E. W. Plummer$^1$}

\affiliation{$^1$Department of Physics $\&$ Astronomy, Louisiana
State University, Baton Rouge, Louisiana 70802, USA}
\affiliation{$^2$Department of Physics $\&$ Texas Center for
Superconductivity, University of Houston, Houston, Texas
77204-5002, USA} \affiliation{$^3$Materials Science $\&$
Technology Division, Oak Ridge National Laboratory, Oak Ridge,
Tennessee 37831, USA} \affiliation{$^4$Department of Materials
Science $\&$ Engineering, The University of Tennessee, Knoxville,
Tennessee 37996, USA}

\begin{abstract}
High-resolution scanning tunneling microscopy (STM) measurements
on BaFe$_2$As$_2$ $-$ one of the parent compounds of the
iron-based superconductors $-$ reveals a (1$\times$1)
As-terminated unit cell on the (001) surface.  However, there are
significant differences of the surface unit cell compared to the
bulk: only one of the two As atoms in the unit cell is imaged and
domain walls between different (1$\times$1) regions display a
$\it{C}$$_2$ symmetry at the surface. It should have been
$\it{C}$$_{2v}$ if the STM image reflected the geometric structure
of the surface or the orthorhombic bulk. The inequivalent As atoms
and the bias dependence of the domain walls indicate that the
origin of the STM image is primarily electronic not geometric. We
argue that the surface electronic topography mirrors the bulk spin
structure of BaFe$_2$As$_2$, via strong orbital-spin coupling.

\end{abstract}

\vskip          15          pt \pacs{68.35.B-, 68.37.Ef, 73.20.-r,
74.40.Xa} \maketitle

The discovery of high-temperature superconductivity in the
Fe-based compounds has generated enormous excitement and activity
in the scientific community \cite{Kamihara, Johnston}. Not only is
this a new class of materials exhibiting some form of
unconventional superconductivity but at the first glance the
behavior resembles that of the cuprates \cite{Andrei}, raising the
expectation that the Fe-based superconductors might offer an
avenue to understand the inherent pairing mechanism responsible
for superconductivity in both systems.  The ground state of the
parent compounds in the cuprates and Fe-based superconductors is
antiferromagnetically (AFM) ordered and it appears that the
magnetic ordering must be suppressed in order to achieve
superconductivity. Both sets of materials exhibit a
superconducting ``dome'' as a function of either hole or electron
doping. However, as more data becomes available for the Fe-based
compounds, it is becoming increasingly clear that the members of
this family behave rather differently from the cuprates. The AFM
ground state of the Fe-based parent compounds is metallic but Mott
insulating for the cuprates.  The small magnetic moments
\cite{Lumsden, Huang} and the characteristic of electronic
structure probed by photoemission measurements \cite{Shimojima}
indicate that the Fe bands are like an itinerant metal not
localized as in the cuprates. While cuprates such as
La$_{2-x}$Sr$_x$CuO$_4$ undergo a structural transition
\cite{Andrei}, there is no evidence for the coupling between
structure and AFM ordering. In Fe-based compounds, there is
complex coupling between lattice and spin degrees of freedom: a
structural transition from a high-temperature tetragonal (HTT) to
a low-temperature orthorhombic (LTO) phase is always accompanied
by a magnetic transition within a narrow temperature window. It is
also known that the application of pressure can drive some of the
parent compounds into the superconducting state without chemical
doping \cite{Simon}. Naively, the creation of a surface can be
viewed as the application of a uniaxial pressure. In this Letter,
we explore the effect on the coupling between spin, lattice and
electrons in one of the parent compounds, BaFe$_2$As$_2$, caused
by the creation of a surface.

Figure 1 shows the bulk and surface structure for the LTO phase of
BaFe$_{2}$As$_{2}$. It consists of alternatively stacking Ba and
Fe-As layers in bulk (Fig. 1a). The LTO phase ($<$ 140 K) is a
collinear AFM ordering with the spin structure shown in Fig. 1a.
We know from our previous study \cite{Nascimento} that the ordered
exposed surface of BaFe$_{2}$As$_{2}$ is the As plane and there is
no measurable surface reconstruction. If the As atoms were buckled
vertically, it would be detected by low energy electron
diffraction (LEED) \cite{Nascimento2}. On the surface, the As
atoms (blue) are in the first plane and the Fe atoms (red) in the
second plane for the (1$\times$1) (001) surface unit cell with the
bulk orthorhombic structure (see Fig. 1b). As shown in Fig. 1b,
all As atoms are expected to be identical in the surface unit cell
exhibiting $\it{C}$$_{2v}$ symmetry. The mystery is that the
high-resolution scanning tunnelling microscopy (STM) image measure
of electronic density distribution, only reveals half of As atoms
that should be present for a bulk truncated surface (Fig. 1c)
\cite{Nascimento}. This suggests that there are two inequivalent
As sites on the surface not seen in bulk. According to our STM
work reported in this Letter, it is plausible that the two
inequivalent As sites result from the underneath spin structure
through strong orbital-spin coupling. Given the fact that the Fe
moments are aligned antiferromagnetically along the longer
$\it{a}$ axis and ferromagnetically along the shorter $\it{b}$
axis \cite{Lumsden}, the relationship between the ``visible" As
atoms and the spin structure is illustrated in Figs. 1d (As2) and
1e (As1). While we cannot determine which of these two
configurations has the lowest energy, As1 and As2 are clearly
surrounded by different spin environments.

\begin{figure}[t]
\includegraphics[width=0.5\textwidth]{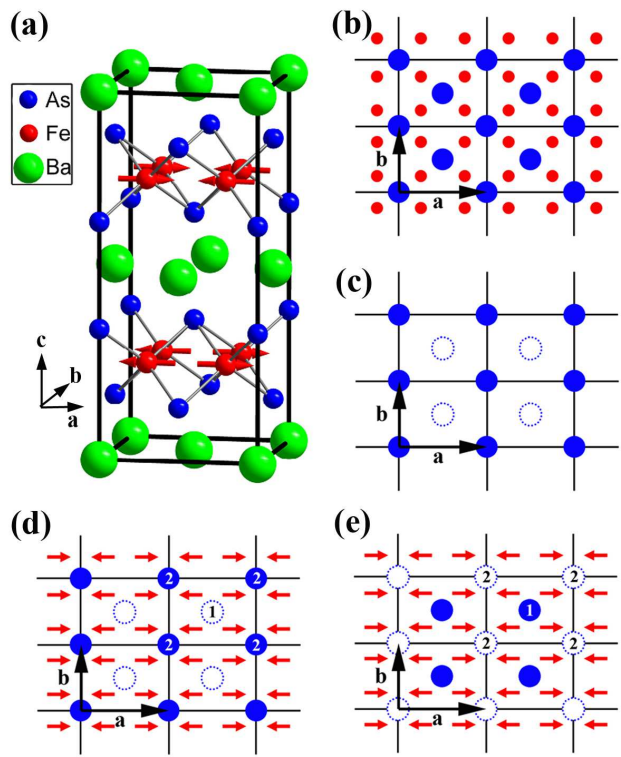}
\caption{(a) Bulk lattice and spin structures of
BaFe$_{2}$As$_{2}$ with Fe magnetic moments indicated by red
arrows; (b) Schematic view of As terminated surface with
underneath Fe layer; (c) As atoms seen by STM (solid circles), the
empty circles represent ``invisible" As atoms; (d-e) Possible
relationship between ``visible" (As2 in (d), As1 in (e)) and the
spin structure of Fe atoms}
\end{figure}

For our STM investigation, we use high-quality BaFe$_2$As$_2$
single crystals that were grown using self-flux method
\cite{Sefat}. The measurements were conducted on a home-built
variable temperature STM with a tungsten tip. Single crystalline
BaFe$_2$As$_2$  was firstly pre-cooled to 80 K in an ultra-high
vacuum environment with basic pressure lower than 5$\times
$10$^{-11}$  Torr. After the $\it{in}$-$\it{situ}$ cleavage, the
sample was immediately inserted into the pre-cooled STM head. Fig.
2a displays a typical STM topographic image with atomic
resolution. In addition to the (1$\times$1) surface structure,
there are white spots either forming zigzag lines (small and
clear) or randomly distributed (large and fuzzy). The large and
fuzzy white spots can be manipulated by the tip, which are likely
Ba atoms as discussed previously \cite{Nascimento}. Similarly,
there are dark spots, some of which are randomly distributed and
the others are aligned with white spots in the zigzag lines.
Importantly, there are always ``dark" spots (see Figs. 2a, 2c)
(impurities/defects/``invisible" As sites) at every corner,
wherever the zigzag lines change the direction. Large-scale
topographic images (not shown) prove that the zigzag lines form
closed loops. Fig. 2b shows that the periodicity is the
(1$\times$1) surface structure of orthorhombic bulk, except that
we only see half of the As atoms in the surface plane. Through
careful calibration including possible thermal drift-induced
error, piezo scanner asymmetry and hysteresis by using
Fourier-Transform(FT)-STM (see the inset of Fig. 2b), we are able
to identify $\it{a}$ and $\it{b}$ directions of the orthorhombic
unit cell which are labelled in Fig. 2b.

\begin{figure}[t]
\includegraphics[width=0.5\textwidth]{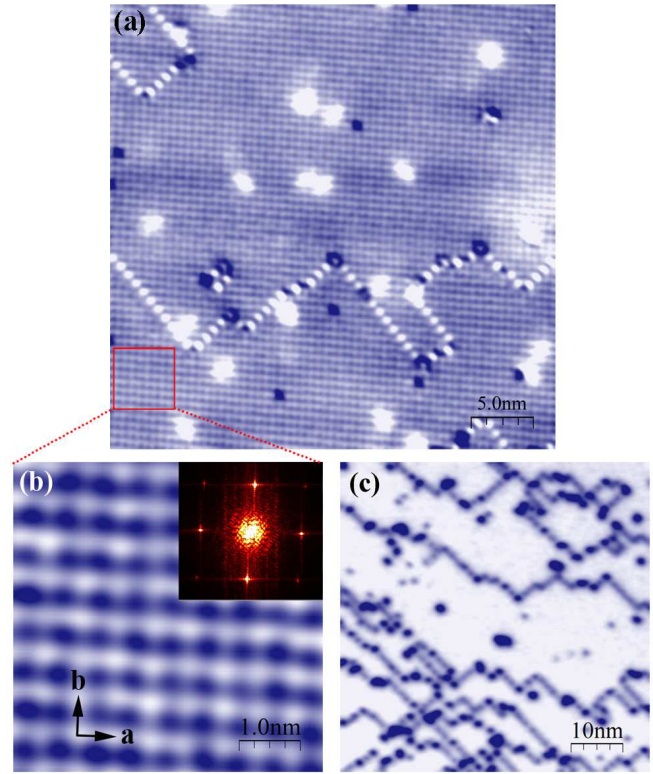}
\caption{(a) A 355 \AA $ $ $\times$ 355 \AA $ $ low-bias
constant-current STM topography (V$_{bias}$ = 23 mV, I$_{tip}$ =
200 pA) on (001) surface BaFe$_{2}$As$_{2}$ at 80 K; (b) An
enlarged 50 \AA $ $ $\times$ 50 \AA $ $ topography from the box in
(a). The $\it{a}$- and $\it{b}$-axis are identified by FT-STM (see
the text); (c) A 700 \AA $ $ $\times$ 700 \AA $ $ high-bias
constant-current STM topography (V$_{bias}$ = 483 mV, I$_{tip}$ =
200 pA).}
\end{figure}

Given the scenario outlined above (strong orbital-spin coupling),
we examine in detail the zigzag lines $-$ domain walls. In solids,
the origin of domains can be geometric, magnetic or electronic.
Using the FT-STM, we find that all domains in the image have the
same $\it{a}$ and $\it{b}$ directions so there is no rotation of
the lattice when crossing a domain wall. This allows us to exclude
the possibility that the domains are caused by a structural
misorientations in the bulk. What we see are surface electronic
domains, which can be verified by looking at the bias dependence.
Fig. 2c shows the topography taken with a high positive bias
voltage (483 mV). Compared to that taken at 23 mV (Fig. 2a), the
domain boundaries have switched from bright to dark. However, we
do not see the switching of ``visible/invisible" As atoms when
changing polarity in the low bias region, as predict by the theory
\cite{Taoxiang}. Because the atomic resolution is lost, whether
the switching occurs at high bias is unknown. It should also be
mentioned that Fig. 2a can be reproduced after changing the bias
voltage from 483 mV back to 23 mV. Therefore the features seen in
Fig. 2c are not due to the change of tip or sample condition. The
domain boundaries seen by STM are primarily electronic in origin.

Figures 3a and 3b show two different domains with two boundaries
in each. There are two boundaries with one along
$\sim$45$^{\circ}$ and another along $\sim$-45$^{\circ}$ (Fig. 3a)
or $\sim$135$^{\circ}$ (Fig. 3b) with respect to $\it {a}$
direction. As can be seen in either figure the change in the
(1$\times$1) domains, when crossing a boundary, is an inversion of
dark to bright in As, i.e., there is a half electronic unit cell
shift across the boundary lines as indicated by lines with arrows
in Figs. 3a and 3b. Such an inversion would not occur if the
boundary is created simply due to the crystal structural
dislocation with half unit cell shift. In our picture, there is no
structural change across this boundary and the boundary is
actually a spin domain wall. While all the bright white spots
residing on both boundaries have elliptical shape, a closer
examination reveals that the white spots along the -45$^{\circ}$
direction are more rounded and the ones in 45$^{\circ}$ direction
more elongated. If one examines the symmetry carefully it is clear
that the domains exhibit $\it{C}$$_2$ symmetry. Rotating the image
in Fig. 3b by 180$^{\circ}$ transforms the 135$^{\circ}$ boundary
into a -45$^{\circ}$ boundary which is identical to the
-45$^{\circ}$ boundary in Fig. 3a, as expected if the boundary
direction is unchanged. If we further reflect the rotated image
about the $\it{a}$ axis, what was the -45$^{\circ}$ boundary
becomes a 45$^{\circ}$ boundary, which does not look like the
original 45$^{\circ}$ boundary shown in Fig. 3a. Thus, the
boundary symmetry is $\it{C}$$_2$ not $\it{C}$$_{2v}$ This is
consistent with the theoretical proposal that $\it{C}$$_2$
symmetry is induced by the magnetic ordering \cite{symmetry}.

The line profile Z(x) in Figs. 3c and 3d presents a quantitative
comparison on the boundaries shown in Figs. 3a and 3b,
respectively. While it is expected that Z(x) oscillates with the
same periodicity along both directions, the amplitude for the
spots in 45$^{\circ}$ direction (red lines) is more than double
compared to that along -45$^{\circ}$ direction (blue lines). This
proves that there are two types of boundaries, reflecting the fact
that, when spin configuration is included, the symmetry is reduced
from $\it{C}$$_{2v}$ to $\it{C}$$_2$. Fig. 3e demonstrates how the
spin structure changes across the boundary, assuming that the spin
structure in Fig. 1d produces bright As atoms. All of our
observations indicate that there is an electronic order at the
surface that reflects coupling between orbits and spins. Using
this model, it is natural to explain the half electronic unit cell
shift between adjacent domains by adopting a $\pi$ phase shift of
the spin order along both AFM $\it {a}$ axis and FM $\it {b}$ axis
when crossing boundaries. The spots seen at the boundaries are
enhanced local density of states due to the orbital overlap
between two ``visible" As atoms. As illustrated in Figs. 3f and
3g, the boundary along 45$^{\circ}$ direction results in brighter
spots compared to that along the 135$^{\circ}$ direction, when
taking into account of spin contribution. This is consistent with
our experimental observation (Figs. 3c and 3d). The spin structure
shown in Fig. 3e suggests that there is a spin flip across the
surface boundary creating an anti-phase (Pi-phase) spin domain
wall.

\begin{figure}[t]
\includegraphics[width=0.45\textwidth]{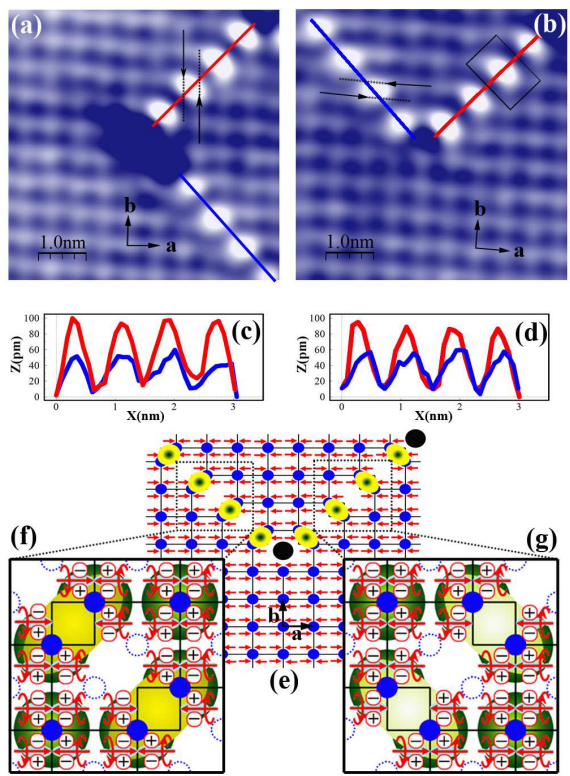}
\caption{ (a-b): Two 56 \AA $ $ $\times$ 56 \AA $ $ low-bias
constant-current STM topographies (V$_{bias}$ = 23 mV, I$_{tip}$ =
200 pA) showing boundary structures at 80 K. The arrows with dash
lines in (a-b) indicate the half electronic unit cell shift in
$\it{a}$ and $\it{b}$ direction respectively. The black
rectangular box in (b) shows the size of bright spot on boundary:
length/width = 3/2 measured from this image; (c-d): the line
profiles for red and blue line in (a) and (b) respectively; (e) A
model for domain walls. Here the blue solid circles denote the
``visible" As atoms in the LTO surface unit cell, and the red
arrows indicate the magnetic moments of Fe atoms; Solid yellow
ellipses represent the bright spots at the boundary; Two black
solid circles represent impurities/defects/``invisible" As sites.
(f-g): Two types of boundaries breaking $\it{C}$$_{2v}$ to
$\it{C}$$_{2}$ symmetry. The light-blue clouds denote the
polarized Fe 3d$_{xz}$ orbitals. Red arrows represent spins and
their orientations. Positive and negative signs in small circles
indicate the phase of orientation. Note that each ellipse along
45$^{\circ}$ direction (g) includes 4 negative signs most close to
white spot (the overlapped two As atoms) by the boundary; while
each along 135$^{\circ}$ boundary direction includes 4 positive
signs (f). }
\end{figure}

The puzzle is why the electronic topography seen with STM mirrors
the spin structure, considering that the Fe layer looks like an
itinerant metal. Recent theoretical \cite{ChiLee, Lv1, Lv2,
Cchen1, Zyin, Daghofer, Cchen2, Frank, Mazin} and experimental
studies \cite{Shimojima, Chu, Akrap} on the electronic structures
of iron-based compounds suggest that orbital degree of freedom
emerges in this multiband system with intimate coupling to
lattice, charge and spin. It was proposed that the ferro-orbital
Fe (3d$_{xz}$) order leads to the structural and magnetic phase
transitions \cite{ChiLee, Lv1}. As a result \cite{Cchen1},
electrical conduction is higher in the AFM $\it{a}$ direction than
that in the FM $\it{b}$ direction as observed experimentally
\cite{Chu}. The recent laser angle-resolved photoemission
spectroscopy and band calculations \cite{Shimojima} indicate that
the two Fermi surface pockets centered at $\Gamma$ point
($\alpha$1 and $\alpha$2) have a predominant Fe 3d$_{xz}$ orbital
component which is polarized by AFM order \cite{Daghofer}. As
argued in Ref. \cite{WeiLee}, most of the detected electronic
contribution by STM comes from the $\Gamma$ centered $\alpha$1 and
$\alpha$2 pockets. In such circumstances, it becomes obvious that
the STM image includes not only the local electronic topography
but also information about magnetic structure.

The first-principles calculation of the surface predicts two
different As surface atoms created by the imbalance in chemical
valances due to missing Ba atoms \cite{Taoxiang}. If this were the
case, one would not expect the difference (intensity) between two
boundaries shown in Fig. 3. Our understanding is that the
calculation did not take into account the impact of magnetic
ordering to the electronic structure. However, there is no doubt
that the surface, especially the As-terminated surface, of
BaFe$_{2}$As$_{2}$ is quite different from the bulk. The key is
how this difference is reflected in the physical properties at or
near the surface, and how the surface may affect the bulk. We have
focused on the As-terminated (1$\times$1) structure because it
corresponds to the bulk orthorhombic phase. But a (1$\times$2)
(tetragonal notation) surface reconstruction \cite{Yin, Chuang,
Massee, Jin, Niestemski, Hui} has also been observed and
associated with the tetragonal bulk structure exhibiting
$\it{C}$$_{2v}$ symmetry. In these materials, these two structures
seem to coexist at the surface \cite{Massee, Jin, Niestemski}
through out the phase diagram of Ba(Fe$_{1-x}$Co$_{x}$)As$_{2}$
\cite{Pan}. Naively, this would indicate that there are patches of
tetragonal surface structure coexisting with orthorhombic phase.
What is needed is a measurement of orthorhombicity as a function
of doping and temperature. It may well be that the (1$\times$1)
surface structure loses its orthorhombicity as a function of
temperature or doping, eventually turnning into a
$\it{C}$($\sqrt{2}$$\times$$\sqrt{2}$)R45$^{\circ}$ surface
reconstruction of a tetragonal bulk. We already know that the
measured superconducting gap using STM is well behaved as a
function of $\it{x}$ in both surface structures \cite{Pan}. It is
possible that the surface ``pins" or ``freezes" the magnetic or
orbital fluctuations resulting in a much higher structural and
magnetic transition temperature than the bulk. If this is true,
the surface may be a nucleation center for the bulk phase
transition.

In summary, the imaged domains and domain walls seen by STM are
shown to be primarily electronic in origin with strong
electron/spin (orbital-spin) coupling, which is clearly reflected
by the $\it{C}$$_2$ symmetry. The intimate coupling between the
spin and electron orbitals at the surface enable us to observe the
electronic structure that mirrors the bulk spin structure. This
offers great opportunities for the investigation on the
orbital-spin coupling. It also opens a new chapter in the
long-standing issues of interplay between superconductivity and
magnetism which may only be present at the surfaces (or under
pressure) of this new class of superconductors.

We would like to thank W. Ku, V. B. Nascimento, P. Phillips, and
D. Singh for fruitful discussion. Research at LSU is partially
supported by NSF DMR-1002622 (GL, RJ, EWP). Research at ORNL is
sponsored by BES Materials Sciences and Engineering Division (ASS,
MAM, BCS, DGM), U. S Department of Energy.


\begin{thebibliography}{34}
\bibitem{Nascimento2} Our LEED system can probe vertical buckling of BaFe$_2$As$_2$ with the sensitivity of 0.05 \AA $ $.
\end{thebibliography}
\end{document}